\documentclass[submission,copyright,creativecommons]{eptcs}
\providecommand{\event}{Infinity 2011}

\usepackage{amssymb}
\usepackage{amsfonts}
\usepackage{amsmath}
\usepackage{theorem}
\usepackage{hyperref}
\usepackage{tikz}
\usepackage{listings}

\newtheorem{thmcounter}{}[section]
\newtheorem{definition}[thmcounter]{Definition}

\usetikzlibrary{arrows,shapes,automata,calc,decorations,decorations.markings,decorations.pathmorphing}
\tikzstyle{every state}=[semithick,node distance=1.5cm,inner sep=1pt,
  minimum size=4.5mm,draw=black,fill=white,initial text=,initial distance=3mm]
\tikzstyle{trans}=[->,semithick,draw=black,>=angle 60]
\tikzstyle{line}=[trans,-]
\tikzstyle{dist}=[circle,draw=black,fill=black,minimum size=0mm,inner sep=1.5pt,node distance=1.75cm]
\tikzstyle{every initial by arrow}=[trans]

\lstnewenvironment{prismcode}[1][]%
{\lstset{%
	morecomment=[l]{//},
	morestring=[b]",
	morekeywords={pta,ipta,module,endmodule,init,clock,invariant,endinvariant,label},
	columns=flexible,
	mathescape=true,
	basicstyle=\small\tt,
	commentstyle=\itshape\color{gray},
	texcl=true,
	frame=single,
	numbers=right,
	numbersep=5pt,
	#1
}}
{}

\lstnewenvironment{xmlcode}[1][]%
{\lstset{%
	language=xml,
	breaklines=true,
	columns=flexible,
	frame=single,
	basicstyle=\small\tt,
	commentstyle=\itshape\color{gray},
	numbers=right,
	numbersep=5pt,
	#1
}}
{}

\newcommand{\halflineup}{\vspace{-0.5\baselineskip}}

\newcommand{\N}{\ensuremath{\mathbb{N}}}
\newcommand{\R}{\ensuremath{\mathbb{R}_+}}
\newcommand{\la}{\langle}
\newcommand{\ra}{\rangle}

\newcommand{\Def}{\overset{\mathrm{def}}{=}}
\newcommand{\trans}[3]{\ensuremath{#1{\,\xrightarrow{#2}\,}#3}}

\newcommand{\true}[0]{\ensuremath{\mathrm{true}}}

\newcommand{\Dist}{\ensuremath{\mathit{Dist}}}
\newcommand{\Adv}{\ensuremath{\mathit{Adv}}}
\newcommand{\IntDist}{\ensuremath{\mathit{IntDist}}}
\newcommand{\I}{\ensuremath{\mathcal{I}}}
\newcommand{\X}{\ensuremath{\mathcal{X}}}
\newcommand{\Z}{\ensuremath{\mathcal{Z}}}
\newcommand{\E}{\ensuremath{\mathcal{E}}}
\newcommand{\A}{\ensuremath{\mathcal{A}}}
\renewcommand{\L}{\ensuremath{\mathcal{L}}}
\renewcommand{\P}{\ensuremath{\mathcal{P}}}
\newcommand{\CC}{\ensuremath{\mathit{CC}}}
\newcommand{\AP}{\ensuremath{\mathit{AP}}}
\newcommand{\inv}{\ensuremath{\mathit{inv}}}
\newcommand{\prob}{\ensuremath{\mathit{prob}}}
\newcommand{\Steps}{\ensuremath{\mathit{Steps}}}
\newcommand{\Support}{\ensuremath{\mathit{Supp}}}
\newcommand{\T}{\ensuremath{\mathcal{T}}}

\newcommand{\finite}{\ensuremath{\mathit{finite}}}
\newcommand{\full}{\ensuremath{\mathit{full}}}

\newcommand{\last}{\ensuremath{\mathit{last}}}
\newcommand{\Paths}{\ensuremath{\mathit{Paths}}}
\newcommand{\Prob}{\ensuremath{\mathit{Prob}}}

\newcommand{\act}[1]{\ensuremath{\mathit{#1}}}

\newcommand{\cc}[1]{\ensuremath{#1}}

\title{Model Checking Probabilistic Real-Time Properties for Service-Oriented Systems with Service Level Agreements}

\def\titlerunning{Model Checking Probabilistic Real-Time Properties for Service-Oriented Systems}

\author{
Christian Krause\footnote{Supported by the 
research school in `Service-Oriented Systems Engineering'
at the Hasso Plattner Institute (HPI).}
\institute{Hasso Plattner Institute\\ 
Prof.-Dr.-Helmert-Str. 2-3\\
D-14482 Potsdam, Germany}
\email{christian.krause@hpi.uni-potsdam.de}
\and
Holger Giese
\institute{Hasso Plattner Institute\\ 
Prof.-Dr.-Helmert-Str. 2-3\\
D-14482 Potsdam, Germany}
\email{holger.giese@hpi.uni-potsdam.de}
}

\def\authorrunning{Christian Krause, Holger Giese}

\begin{document}
\maketitle

\begin{abstract}
The assurance of quality of service 
properties is an important aspect of 
service-oriented software engineering. 
Notations for so-called
\emph{service level agreements} (SLAs), 
such as the Web Service Level Agreement (WSLA) language, 
provide a formal syntax to specify such assurances in terms 
of (legally binding) contracts between a
service provider and a customer.
On the other hand,
formal methods for verification of probabilistic
real-time behavior have reached a level of 
expressiveness and efficiency which allows
to apply them in real-world scenarios. In 
this paper, we suggest to employ the 
recently introduced model of Interval 
Probabilistic Timed Automata (IPTA) for 
formal verification of QoS 
properties of service-oriented systems.
Specifically,
we show that IPTA in contrast to Probabilistic 
Timed Automata (PTA) are able to capture the
guarantees specified in SLAs directly. 
A particular challenge in the analysis of 
IPTA is the fact that their naive semantics
usually yields an infinite set of states
and infinitely-branching transitions. However,
using symbolic representations, IPTA can 
be analyzed rather efficiently. 
We have developed the first 
implementation of an IPTA model checker
by extending the PRISM tool and 
show that model checking IPTA 
is only slightly more expensive than 
model checking comparable PTA.
\end{abstract}


\section{Introduction}\nocite{JHF09,MSKA11}
\noindent
One of the key tasks in engineering service-oriented systems is the assurance of quality of service (QoS) properties, such as `the response time of a service is less than 20ms for at least $95\%$ of the requests'. 
\emph{Service level agreements} (SLAs) provide a notation for specifying such guarantees in terms of (legally binding) contracts between a service provider and a service consumer. A specific example for an SLA notation is the Web Service Level Agreement (WSLA)~\cite{KL03,WSLA} language, which provides a formal syntax to specify such QoS guarantees for web services. The compliance of a service implementation with an SLA is commonly checked at runtime by means of monitoring them. 

However, due to the fact that an application or service may itself make use of other services, guaranteeing probabilistic real-time properties can be difficult. The problem becomes even harder, when the service is not bound to a specific service provider but linked dynamically. Statistical testing of the service consumer together with all currently possible service providers can provide some evidence that the required probabilistic real-time properties hold. However, each time a new service provider is connected or in situations when a known service provider slightly changes the characteristics of the offered service, the test results are no longer representative.
%

In the last couple of years, formal methods for verification of probabilistic real-time behavior have reached a level of expressiveness and efficiency that allows to apply them to real-world case studies in various application domains, including communication and multimedia protocols, randomized distributed algorithms and biological systems (cf.~\cite{PRISMCaseStudies,UPPAALCaseStudies}). Therefore, it is a natural step to investigate also their suitability to address the outlined challenges for guaranteeing QoS properties of service-oriented systems. In particular, dynamically linking of services in service-oriented systems introduces major difficulties concerning the analysis of their QoS properties.

In this paper, we suggest to employ the recently introduced model of Interval Probabilistic Timed Automata~\cite{IPTA} (IPTA) which extend Probabilistic Timed Automata~\cite{KGSS02} (PTA) by permitting to specify intervals, i.e., lower and upper bounds for probabilities, rather than exact values. The contributions of this paper can be summarized as follows:
(1)~We show that IPTA (in contrast to PTA) are able to capture the guarantees specified in SLAs directly. The notion of probabilistic uncertainty in IPTA allows modeling and verifying service-oriented systems with dynamic service binding, where one can rely only on the guarantees stated in the SLA and no knowledge about the actual service implementation is available.
(2)~To the best of our knowledge, we present the first implementation of an IPTA model checker and show that it can analyze IPTA nearly as fast as comparable PTA.
(3)~We show that a naive analysis using sampling of PTA does not yield the correct results as predicted by IPTA. Furthermore, we provide evidence that checking equivalent PTA has a worse performance than checking the IPTA directly.

\subsubsection*{Organization}
The rest of this paper is organized as follows.
Section~\ref{sec:modeling} demonstrates that IPTA
naturally permit to capture the guarantees of an SLA when 
modeling the behavior of a service provider.
Section~\ref{sec:ipta} introduces the syntax and 
semantics of interval probabilistic timed automata.
Section~\ref{sec:model-checking} discusses symbolic
PTCTL model checking and the 
probabilistic reachability problem.
In Section~\ref{sec:tool-support} we present our tool
support.
In Section~\ref{sec:evaluation} we show that IPTA 
checking is only slightly 
more expensive than PTA checking. We show that using 
sampling of the probability values in the intervals to derive a 
representative set of PTA does neither scale as good as IPTA 
checking nor does it work correctly. Finally, we demonstrate that 
also an encoding of IPTA in form of a PTA does not scale as good 
as IPTA checking.
In Section~\ref{sec:related-work} we discuss related work.
Section~\ref{sec:conclusions} contains conclusions
and future work.

\section{Quality of Service Modeling}\label{sec:modeling}
\noindent
Since in the service-oriented paradigm, compositionality is employed to construct new services and applications, the interaction behavior of a service-oriented system can be captured by a set of communicating finite state automata. For instance, a simple service-oriented system can consist of a service provider and a service consumer, both represented as automata, which communicate according to a specific protocol, given by a service contract. 

The QoS of a service-oriented application is often as important as its functional properties. Validation of QoS characteristics usually requires models, which capture probabilistic aspects as well as real-time properties. Probabilistic Timed Automata~\cite{KGSS02} (PTA) are an expressive, compositional model for probabilistic real-time behavior with support for non-determinism. However, a limitation of PTA is the fact that only fixed values for probabilities can be expressed. In practice, it is often only possible to approximate probabilities with guarantees for lower and upper bounds.
%
For this reason, Interval Probabilistic Timed Automata~\cite{IPTA} (IPTA) generalize PTA by allowing to specify intervals of probabilities as opposed to fixed values. This feature is particularly useful to model guarantees for probabilities as commonly found in service level agreements (SLAs).

\begin{xmlcode}[float=htb,
	label={lst:wsla},
	caption={A response time guarantee in WSLA}]
<Metric name="NormalResponsePercentage" type="float" unit="Percentage">
  <Source>ServiceProvider</Source>
  <Function resultType="float" xsi:type="wsla:PercentageLessThanThreshold">
    <Metric>ResponseTime</Metric>
    <Value>
      <LongScalar>20</LongScalar>       <!-- Normal responses take less than 20ms -->
    </Value>
  </Function>
</Metric>

<Obligations>
  <ServiceLevelObjective name="ResponseTimeGuarantee">
    <Obliged>ServiceProvider</Obliged>
    <Expression>
      <Predicate xsi:type="GreaterEqual">
        <SLAParameter>NormalResponsePercentage</SLAParameter>        
        <Value>0.95</Value>             <!-- At least 95% normal responses -->
      </Predicate>
    </Expression>
  </ServiceLevelObjective>
<Obligations>
\end{xmlcode}

As a concrete example of an SLA, Listing~\ref{lst:wsla} contains an adaption of a WSLA specification presented in~\cite{KL03}. In the upper part, a metric called \texttt{NormalResponsePercentage} is defined which contains the percentage of response events which took less than 20ms. 
The actual service level agreement is defined in the lower part in terms of a service provider obligation called \texttt{ResponseTimeGuarantee}. This obligation assures that the percentage of responses that take less than 20ms is at least $95\%$. 

\begin{figure}[thb]
\centering
\input{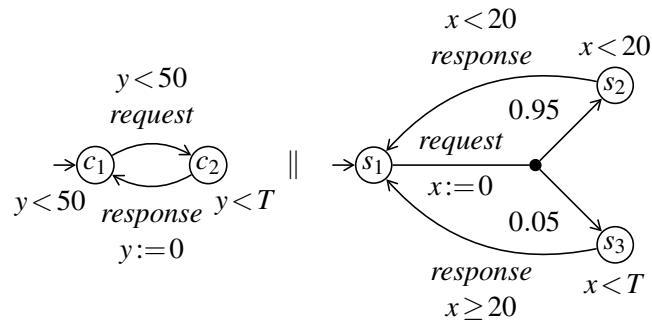}
\vspace{-3mm}
\caption{PTA for a client (left) and a server (right)}
\label{fig:server-pta}
\end{figure}

Figure~\ref{fig:server-pta} depicts a PTA for a client/server application in which the server guarantees a probability of (exactly) 95\% for response times of less than 20ms. The client is modeled as another PTA which synchronizes with the server using the \textit{request} and \textit{response} actions. Note that the client model is actually just a Timed Automata (TA), because no probabilities are employed. However, in the cases where probabilities also matter, we would need exact knowledge of them to be able construct a proper PTA.

This small example shows that PTA, similarly to other automata models, consist of set of states (or \emph{locations}) and transitions (or \emph{edges}).
The time related behavior is specified using clocks (as $x$ in the server) which can be reset ($x := 0$), tested in conditions for transitions ($ x \geq 20$) and also state invariants ($x < 20$ for $s_2$). Note that we use the constant $T$ to denote a constant timeout value in the invariant $x<T$. A clock such as $x$ simply increases with progressing time, unless it is explicitly reset. The conditions block the transition until the clock constraint is fulfilled. Moreover, the state invariant ensures that (1)~no transition leads to this state when this would result in invalidating the state invariant, and (2)~the automaton can no longer stay in this state when this would also lead to a violation of the invariant.
In addition to purely non-deterministic behavior, i.e. when multiple transitions with the same action are enabled, probabilities can be associated with transitions, e.g. $0.95$ for the \textit{request} transition leading to the state $s_2$, and $0.05$ leading to the state $s_3$. Note that for probabilistic transitions, all alternative branches must sum up to $1$. Formally, the target of a transition in a PTA is not a single state, but a discrete probability distribution over the set of all states. Thus, in addition to purely nondeterministic choice, PTA allow to specify the likelihood of an event. Note also that the existence of a parallel operator (written as $\P_1 \parallel \P_2$ where $\P_1$ and $\P_2$ are PTA) moreover allows to synchronize two automata via shared actions, which enables compositional modeling.

\begin{figure}[tbh]
\centering
\input{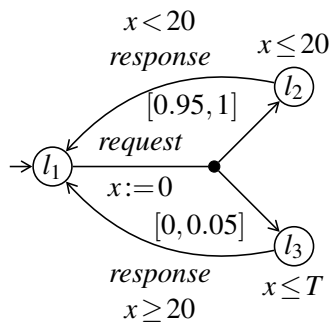}
\caption{IPTA for a simple server}
\label{fig:server-ipta}
\end{figure}

However, the Interval Probabilistic Timed Automaton (IPTA) of a server in Figure~\ref{fig:server-ipta} additionally allows to capture the `at least $95\%$' semantics of the SLA in Listing~\ref{lst:wsla}. The difference to the PTA model is that in IPTA it is possible to specify probabilistic behavior with a level of uncertainty. Specifically, IPTA allow to specify intervals of probabilisties as opposed to the exact probabilities used in PTA. 
The semantics of intervals in contrast to exact values is that each time a probabilistic decision is necessary, any of the usually uncountable many probability distributions which lie within the lower and upper bounds of the intervals denote a valid behavior. 
Therefore, probability intervals match better with the guarantees commonly found in SLAs, such as `with at least 95\% a request is answered within 20ms'.
Note that we did not model the client as another IPTA here, but just as the TA in Figure~\ref{fig:server-pta}. However, similarly to the modeled server, we can also model uncertain probabilistic behavior in the client, such as `with at least 75\% a request is made within 50ms'. The parallel composition of IPTA then allows to derive a model of the complete system. 

Given such models in form of PTA or IPTA, we can now employ model checking to verify probabilistic real-time properties for the composed system, specified in an appropriate probabilistic real-time logic. In our case, we might be interested in the property `the probability that 1 out of 10 responses is too slow is at most 5\%'.
%
%
%
%
%
%
As we will demonstrate later in this paper, there are important differences between the outcome of such an analysis depending on whether we employ the PTA using exact probabilities or the IPTA which allows to specify only lower and upper bounds. In particular, no sample set of PTA derived from the IPTA by choosing values from the interval is in general sufficient to derive the same result as the analysis of the IPTA.

\section{Interval Probabilistic Timed Automata}
\label{sec:ipta}

Interval probabilistic timed automata (IPTA)~\cite{IPTA}
integrate the probabilistic real-time modeling concepts of
probabilistic timed automata (PTA)~\cite{KGSS02} and the 
idea of probabilistic uncertainty
known from interval Markov chains~\cite{SVA06}.
Thus, they not only provide a way to distinguish
between purely probabilistic and nondeterministic 
(timed) behavior, but also allow to specify uncertain
probabilities using lower and upper bounds. These 
ingredients make IPTA a suitable formal model for
the specification and verification 
of QoS assurances that can be commonly
found in SLAs.


\subsection{Preliminaries}\label{sec:prelims}
\noindent
\subsubsection*{Discrete probability distributions}
For a finite set $S$, $\Dist(S)$ is the set of 
\emph{discrete probability distributions} over $S$, i.e.,
the set of all functions $\mu : S \to [0,1]$,
with $\sum_{s\in S} \mu(s) = 1$. 
The \emph{point distribution} $\mu_s^{\bullet}$ is the 
unique distribution on $S$ with $\mu(s)=1$.

\subsubsection*{Clocks, valuations and constraints}
Let $\R$ denote the set of non-negative reals.
Let $\X = \{x_1,\ldots,x_n\}$ be a set of
variables in $\R$, called \emph{clocks}.
An $\X$-\emph{valuation} is a map $v: \X \to \R$.
For a subset $X \subseteq \X$, $v[X:=0]$ denotes the 
valuation $v'$ with $v'(x) = 0$ if $x\in X$ and
$v'(x) = v(x)$ if $x\notin X$.
For $d\in \R$, $v + d$ is the valuation $v''$ 
with $v''(x) = v(x) + d$ for all $x\in \X$.
A \emph{clock constraint} $\zeta$ on $\X$ is an expression 
of the form $x \bowtie c$ or $x-y \bowtie c$ 
such that $x,y\in \X$, $c \in \R$ and
$\bowtie \;\in \{ \leq, <, >, \geq \}$,
or a conjunction of clock constraints.
A clock valuation~$v$ satisfies~$\zeta$, written as 
$v\triangleright \zeta$ if and only if $\zeta$ evaluates
to {\true} when all clocks $x \in \X$ are substituted
with their clock value $v(x)$.
Let $\CC(\X)$ denote the set of all clock constraints 
over $\X$.

\subsection{Syntax}\label{sec:syntax}
\noindent
Before defining IPTA formally, we introduce
a syntactical and, thus, finite notion of 
probability interval distributions.

\begin{definition}[Interval distribution]
\label{def:interval-distribution}
Let $S$ be a finite set.
A \emph{probability interval distribution}
$\lambda$ on $S$ is a pair of functions $\lambda = \la \lambda^\ell,\lambda^u \ra$
with $\lambda^\ell,\lambda^u: S \to [0,1]$, such that
$\lambda^\ell(s) \leq \lambda^u(s)$ for all $s \in S$ and furthermore:
\begin{align}
\label{equ:sumbounds}
\sum_{s\in S} \lambda^\ell(s) \leq 1 \leq \sum_{s\in S} \lambda^u(s)
\end{align}
The set of probability interval distributions over 
$S$ is denoted by $\IntDist(S)$. The \emph{support} of $\lambda$ is 
defined as 
$\Support(\lambda) = \{ s\in S \;|\; \lambda^u(s) > 0\}$.
Let $\lambda^{\bullet}_s$ be the unique interval distribution 
that assigns $\la 1,1 \ra$ to $s$, and $\la 0,0 \ra$ to all $t\in S, t\neq s$.
\end{definition}

A probability interval distribution $\lambda$ is a symbolic representation
of the non-empty, possibly infinite set of probability distributions that 
are conform with the interval bounds:
$\{\, \mu \in \Dist(S) \;|\; \forall s\in S: 
\lambda^\ell(s) \leq \mu(s) \leq \lambda^u(s)\,\}$.
If clear from the context, we may abuse notation and identify 
$\lambda$ with this set and also write $\mu \in \lambda$
if and only if $\mu$ respects the bounds of $\lambda$.
Note that interval distributions are also used 
(in a slightly different syntax) in the
notion of \emph{closed interval specifications} in~\cite{JL91}.
However, the explicit definition using lower and upper interval 
bounds in our model enables a syntactical treatment
of interval distributions which is useful, e.g.,
in the following notion of \emph{minimal interval distributions}.


\begin{definition}[Minimal interval distribution]
\label{def:minimal-interval-distribution}
An interval distribution
$\lambda$
on $S$ is called \emph{minimal} if for all $s\in S$
the following conditions hold:
\begin{enumerate}
  \item \label{itm:minimal1}
  $\lambda^u(s)    + \sum_{t\in S, t\neq s}\lambda^\ell(t) \leq 1$
  \item \label{itm:minimal2}
  $\lambda^\ell(s) + \sum_{t\in S, t\neq s}\lambda^u(t) \geq 1$
\end{enumerate}
\end{definition}

Minimal interval distributions have the property that the bounds of all
intervals can be reached (but not necessarily at the same time). 
Although minimality is formally not needed in the properties that
we consider here, it is often a desirable requirement since it can serve as
a sanity check for a specification. For instance, the 
interval distribution
$\lambda = \{\, s \mapsto \la 0.4,0.5 \ra, t \mapsto \la 0.4,0.5 \ra \,\}$
is not minimal because condition~\ref{itm:minimal2} is violated.
Here, the lower bounds of $0.4$ can never be reached. In fact, the only
probability distribution that is conform with the interval bounds
is $\mu = \{\, s \mapsto 0.5, t \mapsto 0.5 \,\}$. Thus, the minimality 
condition is a useful requirement which allows to verify the validity
of interval bounds. Note also that it is always possible to derive 
a minimal interval distribution from a non-minimal one
by \emph{pruning} the interval bounds, e.g., by setting
$\lambda^u(s) := 1 - \sum_{t\in S, t\neq s}\lambda^\ell(t)$ if 
condition~\ref{itm:minimal1} is violated for the state $s$.


\begin{definition}[Interval probabilistic timed automaton]
An \emph{interval probabilistic timed automaton} is a tuple
$\I = (L,L^0,\A,\X,\inv,\prob,\L)$ consisting of:
\begin{itemize}
  \item a finite set of locations $L$ with $L^0 \subseteq L$ the set of initial locations,
  \item a finite set of action $\A$,
  \item a finite set of clocks $\X$,
  \item a clock invariant assignment function $\inv: L \to \CC(\X)$
  \item a probabilistic edge relation 
		$\prob \; \subseteq L \times \CC(\X) \times \A \times \IntDist(2^\X \times L)$, and
  \item a labeling function $\L: L \to 2^{\AP}$ assigning atomic propositions
to locations.
\end{itemize}
\end{definition}

Note that for more flexibility and a clear separation between 
communication and state invariants, our IPTA model
contains both actions on transitions and atomic propositions
for states. This approach is also in line with our tool 
support based on an extended version of PRISM 
(see Section~\ref{sec:tool-support}).




As an example, we consider the IPTA model
of a simple server depicted in Figure~\ref{fig:server-ipta},
where we denote interval distributions
by small black circles. 
The set of actions is $\A = \{ \mathit{request}, \mathit{response} \}$, 
and the clocks are $\X = \{ x \}$.
For simplicity, we do not include atomic propositions here.
Moreover, we associate the interval 
$[1,1]$ with edges that have a support 
of size $1$.
The server modeled by this IPTA responds to an incoming
request within $20$ms 
with a probability between $95\%$ and $100\%$.
These lower and upper bounds can arise in scenarios
where the exact probabilities are unknown or cannot
be given precisely, e.g., due to implementation details.
For instance, one can imagine that the server relays
all requests to an heterogeneous, internal server 
farm, in which the success probability depends on the
currently chosen server.

\subsubsection*{Composition}

An important aspect of the service-oriented paradigm is
compositionality, i.e., the fact 
that new services can be built by composing existing
ones. Therefore, it is also crucial to support composition
at the modeling level. In our approach, a parallel
operator for IPTA is used for this purpose.
The parallel composition of IPTA
is defined analogously to the one for PTA.
However, we need to compose interval 
distributions instead of probability distributions.

\begin{definition}[Parallel composition]
\label{def:ipta-parallel}
The \emph{parallel composition} of
two interval probabilistic timed automata
$\I_i = (L_i,L^0_i,\A_i,\X_i,\inv_i,\prob_i,\L_i)$ 
with $i\in \{1,2\}$ is defined as:
\begin{equation*}
\I_1 \parallel \I_2 = (L_1 \times L_2, L^0_1 \times L^0_2, \A_1 \cup \A_2,
	\X_1 \cup \X_2, \inv, \prob, \L)
\end{equation*}
such that
\begin{itemize}
  \item $\L(\la l_1,l_2 \ra) = \L_1(l_1) \cup \L_2(l_2)$ 
  		for all $ l_1 \in L_1, l_2 \in L_2$  		
  \item $\inv(\la l_1,l_2 \ra) = \inv_1(l_1) \wedge \inv_2(l_2)$ 
  		for all $ l_1 \in L_1, l_2 \in L_2$
  \item $\la\la l_1,l_2 \ra , \zeta, a, \lambda \ra \in \; \prob$ if and only if one 
  		of the following conditions hold:
  \begin{enumerate}
    \item $a\in \A_1 \setminus \A_2$ and there exists 
    $\la l_1, \zeta, a, \lambda_1 \ra \in \; \prob_1$ such that
    $\lambda = \lambda_1 \otimes \lambda^{\bullet}_{\la \emptyset,l_2 \ra}$
    \item $a\in \A_2 \setminus \A_1$ and there exists 
    $\la l_2, \zeta, a, \lambda_2 \ra \in \; \prob_2$ such that
    $\lambda = \lambda^{\bullet}_{\la \emptyset,l_1 \ra} \otimes \lambda_2$
    \item $a\in \A_1 \cap \A_2$ and there exists 
    $\la l_i, \zeta_i, a, \lambda_i \ra \in \; \prob_i$ such that
    $\lambda = \lambda_1 \otimes \lambda_2$
    and $\zeta = \zeta_1 \wedge \zeta_2$ 
  \end{enumerate} 
\end{itemize}
\hspace{2mm} where for any $l_i \in L_i$, $X_i \subseteq \X_i$:
  \begin{align*}  
  \lambda_1 \otimes \lambda_2 (X_1 \cup X_2, \la l_1, l_2 \ra )^\ell &\Def
  \lambda_1^\ell(X_1,l_1) \cdot \lambda_2^\ell(X_2,l_2)
  \\
  \lambda_1 \otimes \lambda_2 (X_1 \cup X_2, \la l_1, l_2 \ra )^u &\Def
  \lambda_1^u(X_1,l_1) \cdot \lambda_2^u(X_2,l_2) 
  \end{align*}
\end{definition}

Thus, the product of two interval distributions is simply defined
by the product of their lower and upper bounds. Note also that
the parallel composition for IPTA synchronizes transitions via 
shared actions, and interleaves transitions via unshared actions.

\subsection{Semantics}\label{sec:semantics}
\noindent
The semantics of IPTA can be given in terms of
Timed Interval Probabilistic Systems (TIPS)~\cite{IPTA},
which are essentially infinite-state 
Interval Markov Decision Processes (IMDPs)~\cite{SVA06}.

%

\begin{definition}[Timed interval probabilistic system]
A \emph{timed interval probabilistic system} is a tuple
$\T = (S,S^0,\A,\Steps,\L)$
consisting of:
\begin{itemize} 
  \item a set of states $S$ with $S^0 \subseteq S$ the set of initial states,
  \item a set of actions $\A$, such that $\A \cap \R = \emptyset$,
  \item a transition function $\Steps: S \to 2^{(\A \cup \R) \times \IntDist(S)}$,
	such that, if $(a,\lambda) \in \Steps(s)$ and $a\in \R$,
	then $\lambda$ is a point interval distribution, and
  \item a labeling function $\L: S \to 2^{\AP}$ assigning atomic propositions
to states.
\end{itemize}
\end{definition}

\noindent
The operational semantics of a timed interval probabilistic system
can be understood as follows. 
A \emph{probabilistic transition}, written as $\trans{s}{a,\lambda,\mu}{s'}$,
is made from a state $s\in S$ by:
\begin{enumerate}
  \item nondeterministically selecting 
		an action/duration and interval distribution pair $(a,\lambda) \in \Steps(s)$,
  \item nondeterministically choosing a probability distribution $\mu \in \lambda$,
  \item making a probabilistic choice of target state $s'$ according to $\mu$.
\end{enumerate}
A \emph{path} of a timed interval probabilistic system is a non-empty finite 
or infinite sequence of probabilistic transitions:
\[
\omega = s_0 
   \,\xrightarrow{a_0,\lambda_0,\mu_0}\, s_1
   \,\xrightarrow{a_1,\lambda_1,\mu_1}\, s_2
   \,\xrightarrow{a_2,\lambda_2,\mu_2}\, \ldots
\]
where for all $i\in\N$ it holds that 
$s_i\in S$, $(a_i,\lambda_i) \in \Steps(s_i)$, 
$\mu_i \in \lambda_i$ and $\mu_i(s_i) > 0$.
We denote with $\omega(i)$ the $(i+1)$th state of 
$\omega$, and with $\last(\omega)$ the last state of 
$\omega$, if it is finite.
An \emph{adversary} is a particular resolution of 
the nondeterminism in a timed interval probabilistic system $\T$. 
Formally, an adversary $A$ for $\T$
is a function mapping every finite path $\omega$ of 
$\T$ to a triple $(a,\lambda,\mu)$, such that 
$(a,\lambda) \in \Steps(\last(\omega))$ and $\mu \in \lambda$.
We restrict ourselves to 
\emph{time-divergent} adversaries, i.e., we require that
time has to advance beyond any given time bound. This is 
a common restriction in real-time models to rule out 
unrealizable behavior. The set of all time-divergent
adversaries of $\T$ is denoted by $\Adv_{\T}$.

For any $s\in S$ and adversary $A \in \Adv_\T$, we let
$\Paths^A_\finite(s)$ and $\Paths^A_\full(s)$ be the sets of 
all finite and infinite paths starting in $s$ that 
correspond to $A$, respectively.
Under a given adversary, the behavior of a timed interval probabilistic 
system is purely probabilistic. 
Formally, an adversary for a timed interval probabilistic system
induces an infinite discrete-time Markov chain and, thus, 
a probability measure $\Prob^A_s$ over the set of paths 
$\Paths^A_\full(s)$ (cf.~\cite{KSK76} for details).
The semantics of an IPTA can be 
given by a TIPS as follows.

\begin{definition}[TIPS semantics] 
\label{def:tips-semantics}
Given an IPTA
$\I = (L,L^0,\A,\X,\inv,\prob,\L)$.
The \emph{TIPS semantics} of~$\I$
is the timed interval probabilistic system 
$\T_\I = (S,S^0,\A,\Steps,\L')$ where:
\begin{itemize}
  \item $S\subseteq L \times \R^\X$, such that $\la l,v \ra \in S$ if and only if
  $v \triangleright \inv(l)$,
  \item $S_0 = \{\, \la l, v[\X:=0]\ra \;|\; l \in L^0 \,\}$
  \item $\la a,\lambda \ra \in \Steps(\la l,v \ra)$ if and only if one of the
  following conditions holds:
	\begin{itemize}
  		\item Time transitions: 
  			$a=t \in \R$, $\lambda = \lambda^{\bullet}_{\la l,v+t \ra}$ 
  			and $v+t' \triangleright \inv(l)$ for all $0\leq t' \leq t$
  		\item Discrete transitions: 
  			$a\in\A$ and $\la l, \zeta, \hat{\lambda} \ra \in \prob$ 
  			such that $v\triangleright \zeta$ and for any 
  			$\la l',v' \ra\in S$:
  			\begin{itemize}
  			  \item $\lambda^\ell(l',v') = \sum_{X\subseteq \X \wedge v'=v[X:=0]} \hat{\lambda}^\ell(X,l')$
  			  \item $\lambda^u(l',v') = \sum_{X\subseteq \X \wedge v'=v[X:=0]} \hat{\lambda}^u(X,l')$
  			\end{itemize}
	\end{itemize}
  \item $\L'(\la l,v \ra) = \L(l)$ for all $\la l,v \ra \in S$.
\end{itemize}
\end{definition}

%
%
%
%

\section{Symbolic model checking}\label{sec:model-checking}
\noindent
In this section, we recall the symbolic approach 
for PTCTL model checking as introduced for 
PTA in~\cite{KNSW04} and adapted for IPTA in~\cite{IPTA}. 
Moreover, we discuss in more detail an iterative algorithm
for computing the maximum and minimum probabilities for
reaching a set of target states.

\subsection{PTCTL -- Probabilistic Timed Computation Tree Logic}
\noindent
Probabilistic Timed Computation Tree Logic (PTCTL)~\cite{KGSS02}
can be used to specify combined probabilistic and timed 
properties.
Constraints for probabilities in PTCTL 
are specified using the probabilistic 
threshold operator known from PCTL. 
Timing constraints in PTCTL are expressed using
a set of \emph{system clocks} $\X$, which are the clocks
from the automaton to be checked, and
a set of \emph{formula clocks} $\Z$, 
which is disjoint from $\X$.
The syntax of PTCTL is given by:
\begin{align*}
\phi ::= a \;|\; \zeta \;|\; \neg \phi \;|\; 
         \phi \vee \phi \;|\; z.\phi \;|\; 
         \P_{\sim\kappa}[\phi \; \mathcal{U}\, \phi]
\end{align*}
where:
\begin{itemize}
  \item $a\in \AP$ is an atomic proposition,
  \item $\zeta \in \CC(\X \bigcup \Z)$ is a clock constraint over all system and formula clocks,
  \item $z.\phi$ with $z\in \Z$ is a reset quantifier, and
  \item $\P_{\sim\kappa}[\_]$ is a probabilistic quantifier 
		with $\sim \;\in \{ \leq, <, >,$ $\geq\}$
		and $\kappa\in [0,1]$ a probability threshold.
\end{itemize}

As an example for the specification of a combined probabilistic 
and timed property, the requirement 
for a bounded response time, e.g. 
`with a probability of at least 95\% a 
response is sent within 20ms' 
can be formalized in PTCTL as the formula:
\[
z. \P_{\geq 0.95} [ \true \; \mathcal{U} \, 
(\mathit{responseSent} \wedge z< 20)]
\]
Furthermore, it is possible to specify properties
over system clocks, e.g. the formula:
\[
\P_{\leq 0.05}[(x\geq 4) \mathcal{U} (z=8)]
\]
represents the property 
`with a probability of at most 5\%, the system clock 
$x$ exceeds 4 before 8 time units elapse'.
For the formal semantics of PTCTL, 
we refer to~\cite{KGSS02}.


\subsection{Symbolic states}
\noindent
Since the timed interval probabilistic systems
that are being generated as the semantics of an IPTA 
are in general infinite, it is crucial
to find a finite representation which can be used
for model checking. For this purpose, symbolic states 
are considered in~\cite{KNSW04,IPTA}, which are formally 
given by a pair $(l,\zeta)$ of a location $l$ and a 
clock constraint $\zeta$, also referred
to as \emph{zone} in this context.
A symbolic state $(l,\zeta)$ is a finite representation of
the set of state and formula clock valuations
$\{\; \la \la l,v \ra, \E \ra \;|\; v,\E \triangleright \zeta \;\}$.
Based on this finite representation using the notion of zones, PTCTL
model checking is realized by recursively evaluating the parse 
tree of a given formula, computing the set of reachable symbolic 
states.


\subsection{Probabilistic reachability}
\noindent
The probabilistic quantifier $\P_{\sim\kappa}[\_]$ 
can be evaluated
by (i) computing the minimum and maximum probabilities 
for reaching a set of states, which is
also referred to as the problem of 
\emph{probabilistic reachability}, and (ii)
comparing these probabilities with $\kappa$~\cite{KNSW04}.
Formally, the problem of probabilistic reachability
can be stated as follows.
Let $A$ be an adversary for a TIPS $\T = (S,s_0,\A,\Steps,\L)$, 
and $F \subseteq S$ be a set of target states. 
The probability of reaching $F$ from a state $s\in S$ is defined as:
\begin{equation*}
p^A_s(F) = \Prob^A_s \{ \omega \in \Paths_\full^A(s) \;|\; \exists i\in \N: \omega(i) \in F\}
\end{equation*}


\noindent
Then, the \emph{minimal and maximal reachability probabilities} 
of $F$ are defined as:
\begin{equation*}
\label{equ:pminmax}
p^{\min}(F) = \inf_{A \in \Adv_\T} p^A_{s_0}(F) \qquad
p^{\max}(F) = \sup_{A \in \Adv_\T} p^A_{s_0}(F)
\end{equation*}

\subsubsection*{Iterative algorithm}
\label{sec:value-iteration}

The minimum and maximum probabilities 
for a set of target states in a TIPS 
can be computed using
an iterative algorithm \cite{SVA06,IPTA}
known as \emph{value iteration}, which is used
to solve the \emph{stochastic shortest path problem}~\cite{BT91} 
for (interval) Markov decision processes.

Let $\T = (S,S^0,\A,\Steps,\L)$ be a timed interval 
probabilistic system and $F \subseteq S$ be a set 
of target states. Moreover, let $\overline{F} \subseteq S$ 
be the set of states from which $F$ cannot be reached.
We define $(p_n)_{n\in\N}$ as the sequence of probability 
vectors over $S$, such that for any $s \in S$:
\begin{itemize}
  \item $p_n(s)=1$ if $s \in F$ for all $n\in \N$,
  \item $p_n(s)=0$ if $s \in \overline{F}$ for all $n\in \N$,
  \item $p_n(s)$ is computed iteratively if $s \in S\setminus (F\cup \overline{F})$ by:
\begin{align*}
p_0(s) &= 0 \\
p_{n+1}(s) &= \max_{(a,\lambda) \in \Steps(s)} \sum_{t\in \Support(\lambda)} \mu^{\max}_{\lambda}(t) \cdot p_n(t)
\end{align*}
where we consider an ordering $t_1,t_2,\ldots t_N$ of the
states $\Support(\lambda)$, such that the vector
$p_n(t_1), p_n(t_2), \ldots, p_n(t_N)$ is 
in descending order, and $\mu^{\max}_{\lambda}$ is defined
as follows with $m\in \{1,\ldots,N \}$:\footnote{Note that $\sum_{i=k}^m x \Def 0$ whenever $k>m$.}
\end{itemize}
\halflineup
\begin{align*}
\mu^{\max}_{\lambda}(t_m) &= \min\left( \lambda^u(t_m), \left(
            1 - 
            \sum_{i=1}^{m-1}
            \mu^{\max}_{\lambda}(t_i) - \!\!\!
            \sum_{i=m+1}^{N} \!\!
            \lambda^\ell(t_i) \right)\right)
\end{align*}
Then $p_n(s_0)$ converges to $p^{\max}(F)$ for $n\to \infty$. For a correctness proof
of this algorithm we refer to~\cite{IPTA}. Note also that except for the additional
sorting of the support set, the complexity for computing the maximum
and minimum probabilities for IPTA is the same as for PTA.

%
%

Note that PTCTL model checking (interval) probabilistic timed automata
is EXPTIME-complete. However, for certain subclasses of PTCTL
the model checking problem can be shown to be PTIME-complete 
(cf.~\cite{JSL08}).

\section{Tool Support}\label{sec:tool-support}
\noindent
PRISM 4.0~\cite{KNP11} is the latest version of the 
probabilistic model checker developed at the 
University of Oxford. 
For various probabilistic models, including PTA,
PRISM provides verification methods based
on explicit and symbolic model checking,
and discrete-event simulation.

We have extended PRISM~4.0 with support 
for IPTA.\footnote{Our IPTA extension of PRISM is available at
\url{www.mdelab.org/?p=50}.}
Our implementation adds the new operator `$\sim$' to the
PRISM language which can be used to specify probability intervals 
($l \sim u: \dots$) and not only exact probabilities ($0.95: \dots$).
Moreover, we adapted the implementation for 
computing the minimum and maximum probabilities
for reaching a set of target states based on 
the definitions in Section~\ref{sec:value-iteration}.

Listing~\ref{lst:server} contains the PRISM code for the
server IPTA in Figure~\ref{fig:server-ipta} and an IPTA
for a client which performs a fixed number of requests
and then terminates.
The constants \texttt{L} and \texttt{U} are used to declare the 
lower and upper interval bounds for a successful request, e.g. 
by setting \texttt{L=0.95} and \texttt{U=1} we obtain the IPTA
in Figure~\ref{fig:server-ipta}.
Note that we need to set the module type to \texttt{ipta}
to be able to specify probability intervals.
Fixed probabilities are also supported and interpreted as
point intervals. Thus, any PTA model is also a valid IPTA
model in our tool. Note also that the \texttt{invariant} 
section is used in PRISM~4.0 to associate clock 
invariants to locations, such as $x\leq 20$ for the state $s=1$.

\begin{prismcode}[float=t,
	label={lst:server},
	caption={Client/Server system as a PRISM--IPTA}]
ipta

const double L;  // Lower probability for normal response
const double U;  // Upper probability for normal response      
const int REQUESTS; // Number of requests
const int TIMEOUT = 30000; // Timeout value

module Server
  s : [0..2] init 0;
  w : [0..REQUESTS] init 0; // Number of slow responses
  x : clock;
  invariant
    (s=0 $\Rightarrow$ x$\leq$100) & (s=1 $\Rightarrow$ x$\leq$20) & (s=2 $\Rightarrow$ x$\leq$TIMEOUT)
  endinvariant
	
  [request] (s=0 & w<REQUESTS) $\to$ (L$\sim$U):(s'=1)&(x'=0) 
                                   + ((1-U)$\sim$(1-L)):(s'=2)&(w'=w+1)&(x'=0);
  [response] (s=1 & x$\leq$20) | (s=2 & x>20) $\to$ (s'=0)&(x'=0);
endmodule

module Client
  t : [0..REQUESTS] init 0;
  y : clock;
  invariant
	(y<=TIMEOUT)
  endinvariant

  [request] t<REQUESTS $\to$ (t'=t+1)&(y'=0);
  []         t=REQUESTS $\to$ (y'=0);
endmodule

label "lessThan50PercentSlow" = (t=REQUESTS & w<REQUESTS/2);
\end{prismcode}

Note also that we have extended the original server of the example in Figure~\ref{fig:server-ipta} 
here by recording the number of slow responses 
that occurred so far using the variable \texttt{w}.
Moreover, the client now performs only a pre-defined number
of requests, given by the constant \texttt{REQUESTS}. 
This allows us to control and count the number of subsequent 
requests and (slow) responses and to reason about probabilities
for specific scenarios, such as the probability that less than 
50\% of all requests will result in a slow response. 
This particular property is encoded using the label 
\texttt{lessThan50PercentSlow} in line~32.
Note also that this
definition of the client provides a convenient way to scale the size 
of the state space by increasing the number of requests, i.e. the
constant \texttt{REQUESTS}. This is 
particularly useful for conducting benchmarks, e.g. 
for measuring the run-times of the model checker 
for different model sizes (cf. Section~\ref{sec:comparison}).


For the two modules defined in Listing~\ref{lst:server},
PRISM forms the system to be analyzed as 
the parallel composition of the server and the client, 
(cf.~Definition~\ref{def:ipta-parallel}).
In the following section, we give an evaluation of 
our analysis approach and tool support using this example.

\section{Evaluation}\label{sec:evaluation}
\noindent
In this section, we compare the IPTA model in 
Listing~\ref{lst:server} with PTA encodings of
the same example. In particular, we show that
PTA encodings either yield incorrect results
(sampling with exact probabilities) or result in a blow-up of the model which causes
a decay in the run-times of the model checker (equivalent model).

\newcommand{\ipta}{\mathtt{ipta}}
\newcommand{\pta}{\mathtt{pta}}
\newcommand{\sample}{\mathtt{sample}}

\subsection{Difference to sampling}
\begin{figure}[thb]
\centering
\includegraphics[angle=270,width=.5\linewidth]{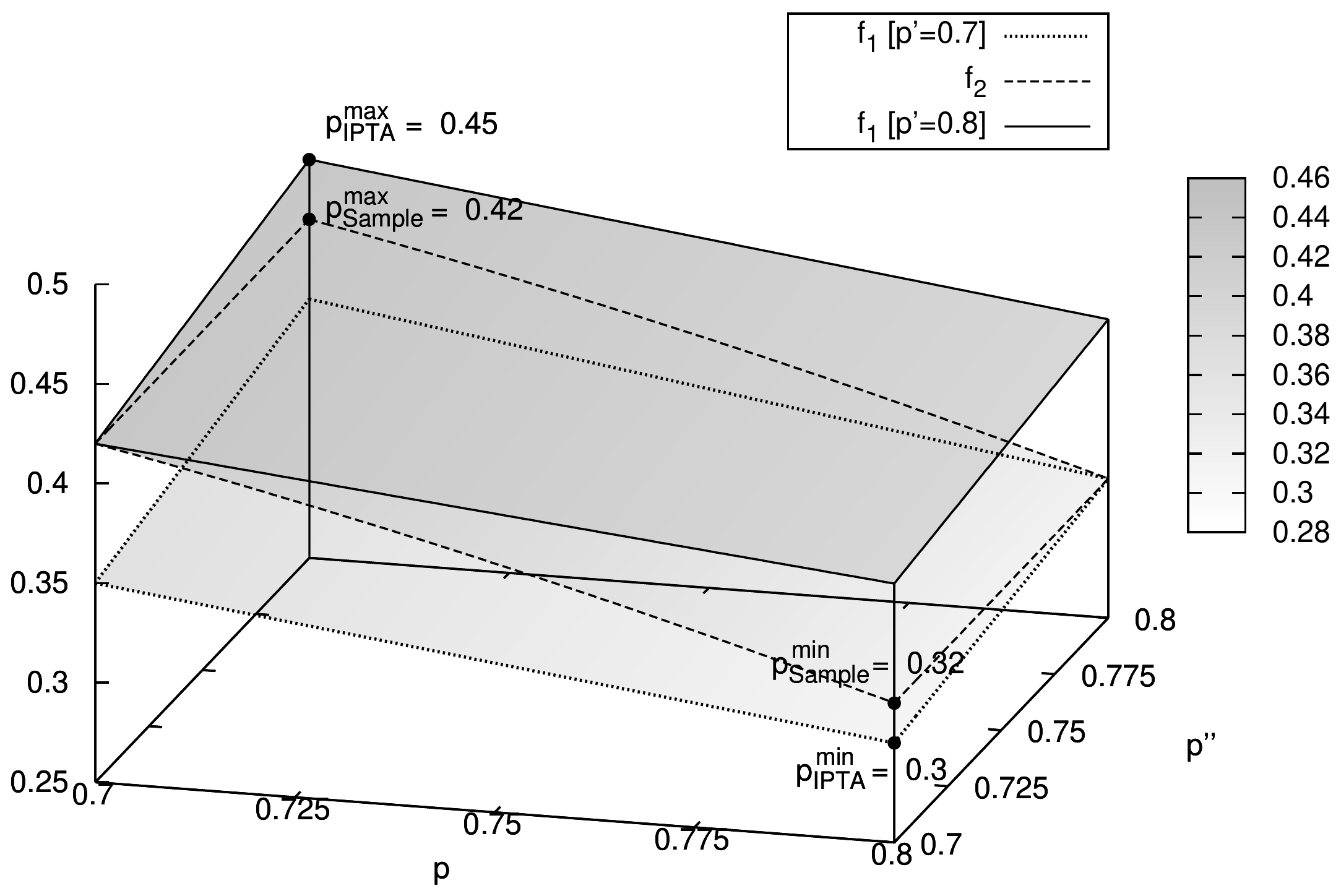}
\caption{Analytic solutions for the property `one out of two response is slow'}
\label{fig:analytic}
\end{figure}

\noindent
For an initial test, we have set the constants in our example to 
\texttt{L=0.7}, \texttt{U=0.8} and \texttt{REQUESTS=2}.
Using the IPTA version of PRISM, 
we then calculated the minimum and maximum 
probabilities for the property that 
one out of two responses was slow: \texttt{(t=2 \& w=1)}.
The computed minimum and maximum probabilities~are:
\[
p_\ipta^{\min} = 0.30,\qquad\qquad p_\ipta^{\max} = 0.45
\]
To illustrate the difference to approaches
with fixed probabilities, we also encoded 
this example as a \texttt{pta} model, where
we tested the following probabilities for 
normal response times:
y=0.7, 0.75 and 0.8. For this model and the above property,
we obtain the following probabilities:
\[
p^{(y=0.7)}_{\pta} = 0.42\qquad
p^{(y=0.75)}_{\pta} = 0.375\qquad
p^{(y=0.8)}_{\pta} = 0.32
\]
It is obvious that these three samples are not 
sufficient to obtain the actual minimum and 
maximum probabilities as predicted using 
the IPTA model. In fact, no
fixed value for $y$ in the interval $[0.7,0.8]$ produces 
the correct results, because the probability
for the chosen property is 
minimal / maximal when $y$ is chosen 
differently for each request. To illustrate this
situation we computed the solutions analytically,
depicted in the graph in Figure~\ref{fig:analytic}.

The plane in the middle represents the solution for
the sampling-based \texttt{pta} approach, which 
reaches a minimum probability of
0.32 for y=0.7 and a maximum probability 
of 0.42 for y=0.8. The
upper and lower plane depict the IPTA version which reaches
a minimum and maximum probabilities of 0.3 and 0.45, respectively.
Therefore, the sampling approach using PTA is not sufficient 
for determining the correct minimum and maximum probabilities
in the original IPTA model.

\subsection{Encoding IPTA as PTA}
\noindent
Although the semantics of an interval distribution, i.e., 
the set of all probability distributions that respect the 
bounds of its intervals, is in general infinite,
it is still possible to encode any finite IPTA 
into an equivalent, finite PTA. 
%
%
%
This encoding, which we also
refer to as PTA$\!^{*}$, works as 
follows:\footnote{The PTA$\!^{*}$ encoding is similar to the MDP reduction of IMDPs in~\cite{SVA06}.}
\begin{itemize}
  \item The actions, clocks and locations of the PTA are the same as in the IPTA.
  \item For every transition $\trans{s}{a}{\lambda}$ in the IPTA
  		and any ordering of the set $\Support(\lambda)$ add
        the transition $\trans{s}{a}{\mu_\lambda^{\max}}$ 
        to the PTA (cf.~Section~\ref{sec:value-iteration}).
\end{itemize}

\begin{figure}[h]
\centering
\halflineup
\input{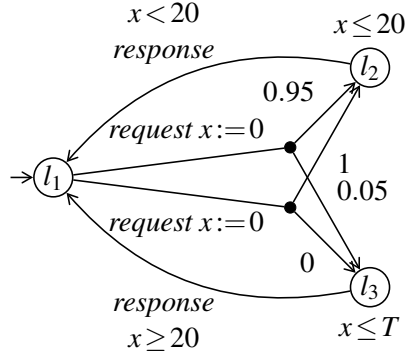}
\halflineup
\caption{PTA$\!^{*}$ encoding of the server IPTA}
\label{fig:encoding}
\end{figure}

As an example, Figure~\ref{fig:encoding} depicts the
PTA$\!^{*}$ encoding of the server IPTA in Figure~\ref{fig:server-ipta}.
From the construction, it is clear that this encoding
preserves probabilistic reachability, i.e., the minimum
and maximum probabilities for reaching a set of target states
in this PTA is the same as for the original IPTA.
However, the number of generated transitions in the PTA
is exponential in the size of the support 
of the transition. Thus, there is a significant blow-up 
in the size of the model. Even in our simple
example in Figure~\ref{fig:server-ipta} where 
the support sets have a size of at most 2, the
larger number of transitions in the PTA$\!^{*}$
encoding results in longer run-times of 
the model checker. To illustrate this, we
increased the number of requests performed
by the client in our running example and compared 
the run-times of PRISM. 

\subsection{Comparison of the run-times}
\label{sec:comparison}

\noindent
Table~\ref{tab:benchmarks} summarizes the 
run-times of our IPTA version of PRISM for
three different encodings of the running example:
\begin{enumerate}
  \item PTA: sampling approach where a single
  probability distribution in the interval distribution is tested;
  \item IPTA: the original model as in Listing~\ref{lst:server};
  \item PTA$\!^{*}$: the encoding of the original IPTA using $\mu_\lambda^{\max}$;
\end{enumerate}

\begin{table}[htb]
\centering
\begin{tabular}{|r|r||r|r|r|}
  \hline
  \#Requests & \#States & PTA & IPTA & PTA$\!^{*}$ \\ \hline \hline
  10 &   235 &  0.752 &  0.804 &  0.816 \\ \hline
  20 &   865 &  2.274 &  2.625 &  2.888 \\ \hline
  30 & 1,895 &  7.274 &  7.818 &  9.225 \\ \hline
  40 & 3,325 & 19.170 & 21.662 & 25.990 \\ \hline
  50 & 5,155 & 43.573 & 47.908 & 57.847 \\ \hline
\end{tabular}
\caption{Runtime in seconds for computing minimum probabilities
for `less than 50\% slow responses'}
\label{tab:benchmarks}
\end{table}

The checking of the PTA version was the fastest. However, 
we have shown above already that such a naive analysis using 
sampling does not produce the correct results. While the 
PTA$\!^{*}$ version yields the correct results, 
the numbers show that the direct checking of the IPTA
is more efficient. This is due to the fact the number
of transitions to be checked in PTA$\!^{*}$ encoding
is higher than in the original IPTA. The actual numbers
of the transitions in the example are listed in 
Table~\ref{tab:transitions}.
Note that in our simple client/server example, the 
support sets of the transitions are very small (of size 1 or 2).
We expect that with a greater branching of transitions, the performance
loss using the PTA$\!^{*}$ encoding gets significantly worse.

\begin{table}[htb]
\centering
\begin{tabular}{|r||r|r|r|}
  \hline
  \#Requests & PTA & IPTA & PTA$\!^{*}$ \\ \hline \hline
10 & 339 &   339 & 521 \\ \hline
20 & 1,269 & 1,269 & 2,031 \\ \hline
30 & 2,799 & 2,799 & 4,541 \\ \hline
40 & 4,929 & 4,929 & 8,051 \\ \hline
50 & 7,659 & 7,659 & 12,561 \\ \hline
\end{tabular}
\caption{Number of transitions for different encodings of the client/server example}
\label{tab:transitions}
\end{table}

\section{Related work}\label{sec:related-work}
\noindent
Probabilistic reachability and expected reachability
for PTA based on an integral model of time 
(digital clocks) is studied in~\cite{KNPS06}.
A zone-based algorithm for symbolic PTCTL~\cite{KNSW04} 
model checking of PTA 
is introduced in~\cite{KNSW04}.
A notion of probabilistic time-abstracting bisimulation for PTA
is introduced in~\cite{CHK08}. 
For an overview of tools that support
verification of (priced) PTA we refer to the 
related tools section 
in~\cite{KNP11}.
%
Interval-based probabilistic models and their
use for specification and refinement / abstraction
have been studied already in '91
in~\cite{JL91}. PCTL model checking of
interval Markov chains is introduced in~\cite{SVA06}.
Symbolic model checking for IPTA is
presented in~\cite{IPTA} based on the approaches
in~\cite{KNSW04,SVA06}. However, no tool support
or evaluation is given. Moreover, we show here
that IPTA can also be encoded into PTA and provide
some empirical data for comparing the differences
in terms of correctness and run-times of our
model checker.

Quality prediction of service compositions based
on probabilistic model checking with PRISM is 
suggested in~\cite{GGMT08}.
A comparison of different QoS models for service-oriented
systems and an extension of the UML for quantitative 
models is given in~\cite{JHF09}. A formal syntax for 
service level agreements of web services can be given
using WSLA~\cite{KL03,WSLA}. A compositional QoS model for
channel-based coordination of services is presented 
in~\cite{MSKA11}.

\section{Conclusions}\label{sec:conclusions}
\noindent
We demonstrated in this paper how the recently introduced model of Interval Probabilistic Timed Automata~\cite{IPTA} (IPTA) can be employed to model and verify quality of service guarantees, specifically, probabilistic real-time properties for service-oriented systems with dynamic service binding with contracts specified in service level agreements. 
We have shown that IPTA can capture the guarantees specified in the SLAs more naturally than PTA.
To the best of our knowledge, our extension of the PRISM tool is the first implementation of an IPTA model checker. Moreover, we were able to show that IPTA can be analyzed nearly as fast as sample PTA and faster than a possible encoding of an IPTA in a finite PTA.

As future work, we plan to study refinement notions for IPTA which we hope will enable us to reason compositionally about QoS guarantees of service-oriented systems. 

%

\subsubsection*{Acknowledgments}
The authors of this paper are grateful
to Dave Parker for his support with
the IPTA implementation in PRISM.

\bigskip

\bibliographystyle{eptcs}
\bibliography{infinity}

\end{document}